%% file: main.tex
\documentclass[acmtog,screen,authorversion]{acmart}   
\acmSubmissionID{papers\_393}  

\usepackage{overpic}
\usepackage{booktabs} 
\usepackage{subcaption,graphicx}

\usepackage{hyperref}
\usepackage[capitalise]{cleveref}  
\crefname{section}{Sec.}{Sec.}
\crefname{table}{Tab.}{Tab.}

\usepackage{comment}
\usepackage{booktabs}
\usepackage{multirow}
\usepackage{mathtools}
\usepackage{cancel}
\usepackage{multirow}
\usepackage{amsmath}
\usepackage{amssymb}
\usepackage{algorithm,algpseudocode}
\DeclareMathOperator*{\argmin}{argmin} 

\usepackage{float}  
\usepackage{stfloats}  
\usepackage{wrapfig}   

\usepackage{color}
\definecolor{blue}{rgb}{0,0,0.6}
\definecolor{green}{rgb}{0,0.6,0}
\definecolor{red}{rgb}{0.6,0,0}
\definecolor{gray}{rgb}{0.4,0.4,0.4}
\definecolor{black}{rgb}{0,0,0}
\definecolor{lightgray}{rgb}{0.83, 0.83, 0.83}
\definecolor{purple}{rgb}{1,0,1}

\usepackage{enumitem}

\newenvironment{tightitemize}{
        \begin{itemize}
                \setlength{\itemsep}{1pt}
                \setlength{\itemindent}{0pt}
                \setlength{\parskip}{0pt}
                \setlength{\parsep}{0pt}
                \setlength{\leftmargin}{0cm}
        }{\end{itemize}}

\AtBeginDocument{%
  }

\setcopyright{rightsretained}
\acmJournal{TOG}
\acmYear{2023} 
\acmVolume{42} 
\acmNumber{6} 
\acmArticle{197} 
\acmMonth{12} 
\acmPrice{15.00}
\acmDOI{10.1145/3618351}

\begin{document}

\title[GarmentCode: Programming Parametric Sewing Patterns]{GarmentCode: Programming Parametric Sewing Patterns}

\author{Maria Korosteleva}
\email{maria.korosteleva@inf.ethz.ch}
\affiliation{
    \institution{ETH Zurich}
    \city{Zurich}
    \country{Switzerland}}
\orcid{0000-0001-7151-0946}

\author{Olga Sorkine-Hornung}
\email{sorkine@inf.ethz.ch}
\affiliation{
    \institution{ETH Zurich}
    \city{Zurich}
    \country{Switzerland}}
\orcid{0000-0002-8089-3974}

\renewcommand{\shortauthors}{Korosteleva and Sorkine-Hornung}

\begin{abstract}
Garment modeling is an essential task of the global apparel industry and a core part of digital human modeling. Realistic representation of garments with valid sewing patterns is key to their accurate digital simulation and eventual fabrication. However, little-to-no computational tools provide support for bridging the gap between high-level construction goals and low-level editing of pattern geometry, e.g., combining or switching garment elements, semantic editing, or design exploration that maintains the validity of a sewing pattern.
We suggest the first DSL for garment modeling -- GarmentCode -- that applies principles of object-oriented programming to garment construction and allows designing sewing patterns in a hierarchical, component-oriented manner. The programming-based paradigm naturally provides unique advantages of component abstraction, algorithmic manipulation, and free-form design parametrization. We additionally support the construction process by automating typical low-level tasks like placing a dart at a desired location.
In our prototype garment configurator, users can manipulate meaningful design parameters and body measurements, while the construction of pattern geometry is handled by garment programs implemented with GarmentCode. Our configurator enables the free exploration of rich design spaces and the creation of garments using interchangeable, parameterized components. We showcase our approach by producing a variety of garment designs and retargeting them to different body shapes using our configurator. The library and garment configurator are available at \href{https://github.com/maria-korosteleva/GarmentCode}{https://github.com/maria-korosteleva/GarmentCode}.

\end{abstract}



\input{fig_tex/teaser.tex}
\maketitle
\input{sec/1_introduction.tex}
\input{sec/2_related_work}
\input{sec/3_method.tex}

\input{sec/4_application_garments.tex}

\input{sec/5_discussion.tex}

\input{sec/6_acks.tex}   

\input{fig_tex/random_design_samples.tex}
\input{fig_tex/code.tex}

\clearpage
\balance
\bibliographystyle{ACM-Reference-Format}
\bibliography{mend_copy,additional}

\input{sec/Supplementary}


\end{document}

%% file: fig_tex/teaser.tex

\begin{teaserfigure}
  \centering
  \begin{overpic}[trim=0cm 0cm 0cm -0.2cm,clip,width=0.999\linewidth,grid=false]{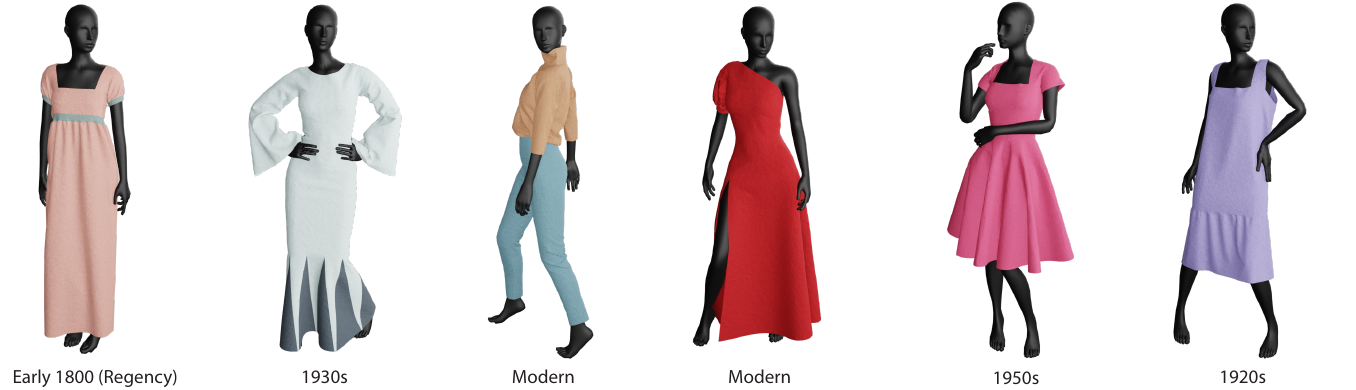}
  \end{overpic}
  \caption{Dress examples showcasing women's garment styles inspired by different epochs in fashion history. All are sampled from a single parametric garment configurator created with GarmentCode. 
  } 
  \label{fig:historical_header}
\end{teaserfigure}

%% file: sec/1_introduction.tex
\section{Introduction}
\label{sec:intro}

``Instead of making people want what we made, we will make what they want''. This motto is one of the motivations behind product configurators and other services that aim to create fashion products tailored to the needs of individual customers. In stark contrast, the fast-fashion industry focuses on mass-production of a variety of designs in standard sizes, relying on trends and statistics of body shapes. The customers are left to choose from available designs and standardized sizes, which often do not fit their body shape. 
This results in customer dissatisfaction, as well as the production of dead stock, which intensifies the negative impact of the garment industry on the climate~\cite{bick2018global}. Unfortunately, the creation of custom-made garments is an expensive, labor-intensive process, unattainable for most people. Recent years saw launches of services for made-to-order clothes of popular garment types, such as Amazon t-shirts~\cite{Amazon:MadeForYou} and Unspun jeans~\cite{Unspun}, offering a middle ground between the two extremes mentioned above. These services allow making some limited adjustments to pre-defined designs and producing garments on demand according to the customer's individual body shape. However, extending such services to general garment designs remains a challenge. Current production-grade tools are oriented towards creating single designs: tools like Clo3D~\cite{clo3d} do not support creating parametric garments. Existing research works on alternative garment modeling tools either do not take sewing patterns into consideration or do not come close to supporting the required complexity, see~\cref{tab:baseline_comparison}.

To facilitate the development of personalized clothing designs, we propose GarmentCode, a domain-specific language (DSL) for sewing pattern construction, adapting the principles of object-oriented programming to naturally allow parameterization and algorithmic support while efficiently handling the design complexity. First, GarmentCode embodies a hierarchical representation of garment sewing patterns with abstract components: the basic component of a garment is a panel, which is a 2D piece of fabric, and more involved, higher-level components can be composed and modified using a range of provided operators. Second, GarmentCode extends stitch definition to allow specification of connectivity between high-level components through semantic component interfaces. Stitch abstraction enables the interchangeability of components with the same semantic interfaces even if the underlying geometry of those differs. In turn, this interchangeability enables modularity: simplicity of integrating novel components into the system and easy garment construction from existing components (see an example in~\cref{sec:eval:complexity}).
Third, our method supports the construction of advanced garment features, such as gathers and darts, currently missing in existing large-scale datasets and modeling approaches~{\cite{Korosteleva2021GeneratingPatterns,Zhu2020,Bertiche2020}}. An example of a garment program is given in~\cref{fig:code}.

Our conceptual framework enables the definition of rich parametric design spaces, which we demonstrate in our garment configurator. It allows the construction of a variety of garment styles, from simple tops, skirts, and pants, to more elaborate complex evening gowns (\cref{fig:historical_header,fig:designsamples}), while using a limited number of parametric components, and with support for adjusting the designs according to different body measurements. A key advantage of using a configurator is the automatic maintenance of a valid sewing pattern and the inherent interchangeability of components, enabling effortless design exploration without having to worry about low-level sewing constraints or deep expert knowledge of patternmaking. Another advantage is the ability to adapt garment designs by exposing intuitive, physically meaningful parameters, such as various body measurements and style parameters.

Limited versions of such configurators can be shared with end customers to let them adjust designs within the limits acceptable for fabrication or to dress custom virtual avatars in video games and metaverses. Such features would welcome users to become part of the creative process and would allow to better satisfy their individual tastes. More detailed design space definitions may assist designers in quickly obtaining starting sewing patterns for their creative exploration. Another envisioned application is the creation of parametric templates for synthetically generated garment design collections, which currently often suffer from limited variety and simplistic designs~\cite{Korosteleva2021GeneratingPatterns,Zhu2020,Bertiche2020}. Such design datasets play an important role in different data-driven applications~{\cite{Wang2018,Jiang2020,Korosteleva2022,Chen2022Structure-PreservingMachines}, attracting much research interest in recent years.

Our implementation of GarmentCode and configurator is publicly available on \href{https://github.com/maria-korosteleva/GarmentCode}{GitHub}\footnote{\href{https://github.com/maria-korosteleva/GarmentCode}{https://github.com/maria-korosteleva/GarmentCode}}.

\input{fig_tex/comparison_table}

%% file: fig_tex/comparison_table.tex
\begin{table}[t]
\small
\caption{Comparison of GarmetCode with baseline systems. Here, ``Definition'' means the ability to specify a single sewing pattern, ``Tools'' refer to instruments supporting pattern construction, ``Modular construction'' enables specifying garment parts as independent modules and constructing new designs by part combinations, ``Continuous'' and ``Categorical'' are parameters with corresponding value types, while a ``Dependent'' parameter's value depends on other parameters, allowing for complex parameterization and resolution of parameter value conflicts. Commercial CAD tools like Clo3D~\shortcite{clo3d} do not support true parameterization while existing parametric garment systems~\cite{Korosteleva2021GeneratingPatterns} support only limited types of parameterization and do not provide any modeling tools; neither of them fully supports modularity.
*\textit{Limited support.} 
}
\begin{tabular}{@{}ccccc@{}}
\toprule
 &
   &
  \begin{tabular}[c]{@{}c@{}}CAD \\ (Clo3D \shortcite{clo3d})\end{tabular} &
  \begin{tabular}[c]{@{}c@{}}Korosteleva \\ and Lee \shortcite{Korosteleva2021GeneratingPatterns}\end{tabular} &
  \begin{tabular}[c]{@{}c@{}}GarmentCode \\ (ours)\end{tabular} \\ \midrule
\multirow{4}{*}{\rotatebox[origin=l]{90}{Modeling\hspace{0.3cm}}}
                          & Definition  
                            & \checkmark & \checkmark & \checkmark \\ \cmidrule(l){2-5} 
                          & Tools
                            & \checkmark & $\times$ & \checkmark* \\ \cmidrule(l){2-5} 
                          
                          & \begin{tabular}[c]{@{}c@{}}Modular \\ construction\end{tabular}    & 
                          \checkmark* & $\times$ & \checkmark  \\ \midrule 
\multirow{3}{*}{\rotatebox[origin=l]{90}{Parameters}} &
  Continuous &
   $\times$ & \checkmark & \checkmark \\ \cmidrule(l){2-5} 
                        & Categorical & 
                        $\times$ & $\times$ & \checkmark \\ \cmidrule(l){2-5} 
                          & Dependent                                                       & $\times$ & $\times$ & \checkmark \\ \bottomrule
\end{tabular}

\label{tab:baseline_comparison}
\end{table}

%% file: sec/2_related_work.tex
\section{Related Work}
\label{sec:rel_work}

\subsection{Garment modeling}
Industry-grade tools for garment modeling, such as Clo3D~\shortcite{clo3d}, rely on artists manually drawing and adjusting sewing pattern shapes. Such tools enable the creation of complex garments, but the design process is often tedious. {Support for semantic parameterization is very limited: Clo3D provides only one pattern parameterized by body measurements -- a bodice, and it is non-extensible with a fixed parameter set. The modular configurator is akin to categorical parameterization for component combination, but it offers a fixed set of component types (upper body, sleeve, collar, cuff) and has only one hierarchy level, and no other parameterizations are supported (\cref{tab:baseline_comparison}).} 
Alternative methods for garment modeling are a subject of ongoing research: for example, editing a garment model in 3D, and automatically readjusting~\cite{Bartle2016c} or inferring a sewing pattern~\cite{Wang2009InteractiveCurves,Meng2012,Liu2018c,Wolff2023DesigningMovement,Pietroni2022} corresponding to the 3D garment design.
Another inspiring line of work aims to maximally reduce the modeling effort by computing 3D garment models from designs sketches~\cite{Li2018FoldSketch:Folds,Wang2018,Fondevilla2021FashionSketches,Chowdhury2022GarmentModeling}, reconstruct sewing patterns from images~\cite{Jeong2015c,Yang2018c}, or 3D capture~\cite{Hasler2007ReverseGarments,Chen2015,Bang2021EstimatingData,Korosteleva2022}.  
However, all these methods focus on producing a single garment. GarmentCode offers an alternative perspective on garment modeling, offering a design toolkit for \emph{parametric} sewing patterns, which allow for fast and convenient exploration of a created design space. Moreover, it offers capabilities to explicitly condition the design on body measurements.
Working in the space of sewing patterns ensures controllable and fabrication-plausible garment designs at each stage of the construction process.

\subsection{Garment modeling at scale}

The rise of deep learning sparked an interest in data-driven techniques in many domains, including various tasks related to garments, such as virtual try-on, modeling, and neural simulation, hence creating a unique demand to generate garment designs at scale to be used in synthetic datasets for training. Some of the works in the area~\cite{Chen2015, Bertiche2020} rely on the combinatorial effect of constructing designs from a set of sub-components, such as multiple options for sleeves, upper body, and lower body garments. In these works, the combinations are performed on 3D geometry, which may result in physically implausible 3D models. These approaches do not provide corresponding sewing patterns. Other works~\cite{Wang2018,Jiang2020,Korosteleva2021GeneratingPatterns} rely on sampling designs from a set of custom parametric {sewing pattern} 
templates, varying the continuous style parameters like length and width of garment elements. \citet{Korosteleva2021GeneratingPatterns} provide a framework for describing garment templates, but do not allow defining discrete parameters, or parameter dependencies, nor  support for base sewing pattern description, {as indicated in~\cref{tab:baseline_comparison}}, rendering it hard to use for complex designs. 

GarmentCode combines continuous and discrete approaches to design variation into one framework, allowing creating garment design templates with interchangeable components and flexible parameterizations of style and body shape. Different helper operators (\cref{sec:arc:helpers}) support the construction process to reduce the workload. 

\subsection{Garment retargeting}
The process of retargeting a garment from one body shape to another is usually performed manually, with leading industry tools like Clo3D~\cite{clo3d} providing support in storing (arbitrary) displacements, specified by designers for each vertex of garment panels individually, for each size. However, a number of research works have been exploring automatic solutions to this problem. Optimization-based approaches~\cite{Brouet2012b,Lee2013c,Lee2018HeuristicCustomization,Fondevilla2021FashionSketches,Wang2018c,AitMouhou2021} transform the garment geometry to reproduce design parameters such as fit, proportionality and overall design shape on a new body model, offering impressive results. In recent years, data-driven approaches for transferring garments across different shapes~\cite{Bertiche2020,Shi2021b,Tiwari2020}, or both shapes and poses~\cite{Ma2020,LalBhatnagar2019,Santesteban2021Self-SupervisedTry-On,Corona_2021_CVPR} have gained popularity. Most of these works represent garments as a displacement map over a body model, which helps disentangle design from body shapes, while~\cite{Corona_2021_CVPR} utilizes an implicit function as a more general approach to describe various garment styles. 
Unlike the approaches mentioned above, GarmentCode embeds the retargeting capability already at the sewing pattern modeling stage, which both reduces the need for manual editing and allows for controllable results reflecting the intention of the designer, which is not offered by the automatic methods mentioned above. The work of \citet{Wang2003FeatureSketches} provides control over the design transfer optimization process by allowing to relate design and body feature points in 3D. 
Some recent works~\cite{Wang2022AnTechniques,Jin2023DesignDesign} present case studies of one or two garment templates implemented with body shape parametrizations. Although they do not propose DSL-like tools to support the implementation of a general garment, these works demonstrate an interest in the fashion field in programmable sewing patterns. Works like \cite{Umetani2011,Yang2018c,Wang2018,Vidaurre2020} build their methods around parameterized base garments. Our work contributes towards a unified framework for creating such designs.

\subsection{Procedural modeling {and CAD DSL}}

Coding shapes like programs is not a novel idea. {There are a number of programming languages developed for traditional solid CAD (Featurescript~\cite{Featurescript}, OpenSCAD~\cite{openscad}
) with rich toolkits supporting shape definition (e.g., collections of standard shapes), editing (e.g., extrusion, boolean operations), and parameterization.} Procedural modeling methods for buildings~\cite{Muller2006ProceduralBuildings,Haegler2010Grammar-basedFacades,Schwarz2015AdvancedArchitecture}, city landscapes~\cite{Parish2001ProceduralCities,Birsak2022Large-ScaleNetworksb} and  plants~\cite{Lindenmayer1968MathematicalDevelopment,Aono1984BotanicalGeneration,Oppenheimer1986RealTrees,Lane2002GeneratingCommunities,Makowski2019} provide tools (e.g., shape grammars) to code highly parametric generative models of the target objects in a variety of styles, with built-in editing options (e.g.,  varying the number of floors and windows in a building). Similar approaches emerge for furniture~\cite{Jones2020ShapeAssembly:Synthesis,Pearl2022GeoCode:Programs}, providing an interesting new angle on reconstruction problems. Due to the unique coupling of the 2D base representation with highly deformable behavior in 3D, garment engineering presents its own challenges and requires targeted modeling tools, but the research in this direction is limited. In addition to works on synthetic garment datasets described above, research on procedurally generated knitting instructions~\cite{Jones2020ComputationalTemplates} computes machine-knittable patterns of given garment models. 
 GarmentCode aims to fill the gap in the procedural generation of sewing pattern designs. We ``translate'' some of the tools of traditional CAD DSL to the garment domain, e.g., our edge loop definition for panels mirrors the structure of parametric boundary representations, and component copy operations are akin to linear and circular patterning in CAD. At the same time, we introduce component abstraction, tools for component stitching, and 2D-3D coupling, unique to garments. 

%% file: sec/3_Method.tex
\input{fig_tex/construction.tex}

\section{Architecture}
\label{sec:architecture}

\input{fig_tex/arc_overview.tex}

\subsection{Overview}

Approaching sewing pattern modeling with a programming-based paradigm, especially when built upon the basis of existing general-purpose programming languages like Python, immediately provides a number of benefits, such as performing auxiliary computations (examples in~\cref{sec:help:projection,app:sec:curved_elements}), free-form parametrization of geometry, and leveraging existing libraries built by the community. None of these benefits are available in existing design representations, be it visual~\cite{clo3d} or text-based parametric approaches~\cite{Korosteleva2021GeneratingPatterns}. However, specifying pattern geometry as a program without the support of structures and tools designed to handle garment-specific properties is tedious and inefficient, akin to the dictionary-based specification of the existing text-based approach~\cite{Korosteleva2021GeneratingPatterns}, requiring explicit definitions of each panel vertex and cross-referencing individual edges for stitch specification (``flat pattern representation''), with no support for element reuse beyond basic language capabilities. 

Hence, the goal of GarmentCode is to provide a domain-specific language for specifying parametric sewing patterns and allowing easy reuse of defined garment elements to compose new garments as modular programs, enabling programming efficiency and complexity management. Specifically, we are bringing the principles of encapsulation and abstraction from the object-oriented programming (OOP) paradigm to garment construction. OOP has proven to be extremely efficient when it comes to building large complex systems across application areas, and we would like to leverage that efficiency when representing the complexity of garment design spaces. A \emph{panel}, a stitched combination of panels, or a higher-level combination of components all define a garment \emph{component} object, which encapsulates its particular geometry and only exposes an abstract semantic \emph{interface}, implemented as a subset of edges of panels that comprise the component. Interfaces of two individual components can be connected together (\emph{abstract stitch}) to form a higher-level component, and any components that implement the same set of interfaces (in our implementation, interfaces identified by the same names) can be used interchangeably regardless of the differences in their encapsulated geometry, enabling modular construction. 

To summarize, the GarmentCode architecture allows pattern description through the following basic types: \emph{Edge}, \emph{EdgeSequence}, \emph{Component}, \emph{Panel} (as a special type of Component), and \emph{Interface}, as illustrated in~\cref{fig:architecture}. Supporting the process, a variety of tools are implemented: a factory for typical edge sequences (e.g., dart shape), specification of curved edges with target properties, projecting an open edge sequence on a corner or an edge, copy operators, normal evaluation for automatic right/wrong side definition, placing stitched components next to each other. To support downstream processing, GarmentCode also implements unfolding the abstract stitch definition into a flat pattern representation and serialization of patterns into files.

Below we describe the elements that comprise our architectural approach in detail.


\subsection{Building blocks}

\subsubsection{Component} 

A component is an abstract class providing a framework to describe a compound garment or a garment element and holds some component processing methods (serialization, rotation, translation, mirroring, {etc.}). Any component should contain the following attributes: 
\begin{tightitemize}
    \item A set of subcomponents;
    \item Stitches -- a list of stitching rules describing how the subcomponents should be connected (see~\cref{sec:arc:stitches}).
    \item A set of interface objects that describe how other components can connect to this one (see~\cref{sec:arc:interfaces}).
\end{tightitemize}
Apart from specifying these attributes, a component construction process may contain instructions for modifying subcomponents, e.g., projecting an interface -- a selection of edges from subcomponent's panels -- of one component onto another, or smart copies, as demonstrated in~\cref{fig:construction}.

\vspace{-1mm}
\subsubsection{Panel} 

A panel is a ``leaf'' component with special structure, so that it can act as a subcomponent, but also specify the panel geometry.
Following the work of~\citet{Korosteleva2021GeneratingPatterns}, GarmentCode defines a panel as a closed piecewise smooth curve represented as a sequence of directed edges organized in a loop, as well as 6D placement parameters (rotation and translation). The latter is needed to correctly place the panel around the body but defaults to zero translation and rotation, and can be left to be set by higher-level components. These attributes define a panel component, in addition to the standard component attributes, namely interfaces and stitches. Note that a panel may contain stitches between its own edges, e.g., if the panel contains darts, as in the fitted bodice panel in~\cref{fig:construction}.

\vspace{-1mm}
\subsubsection{Edges}

An edge is an elementary building block of panels in a sewing pattern. Every edge describes an oriented curve segment. Specifically, GarmentCode supports straight line segments, circular arcs, and quadratic and cubic Bézier splines as edges. This set is flexible and representative enough to model a variety of panel geometries, while ensuring smoothness and computational feasibility. 

Edges are represented by their start and end vertices as attributes (the vertex coordinates are defined in 2D). Bézier curves additionally hold the coordinates of the their control points, while circular arcs store the signed distance of the midpoint of the arc to the straight line connecting the start and end vertices. Building upon the ideas of~\cite{Korosteleva2021GeneratingPatterns}, all controls are specified in a local coordinate system of an edge: the straight segment connecting the edge endpoints is used as the unit horizontal axis, and the left perpendicular is the vertical axis. Such relative representation is invariant to edge translation and rotation, and preserves the curvature with uniform scaling, allowing to perform these operations on all types of edges only though vertex manipulation. 

Working with edges is supported by a variety of routines. For simplicity of use, GarmentCode supports conversion of internal representations from and to standard ones: absolute control point coordinates for Bézier curves, and for circular arcs, the standard three-point representation or the desired radius with flags indicating one of the four arc options. 

\vspace{-1mm}
\subsubsection{Edge sequences} 

An edge sequence specifies an ordered list of edges. An edge sequence used in panel definition must have all its edges chained one after another and into a loop, but other types of edge sequences might be used in other contexts, for example, in interfaces edges may not be chained together nor form a loop. 
To manipulate edge sequences conveniently, GarmentCode implements them as a variation of Python list type and thus supports indexing and slicing, appending, removals, and insertions, as well as domain-specific geometry manipulation methods: translation, rotation, scaling, and reflection around an arbitrary axis.

\vspace{-1mm}
\subsection{Stitches}
\label{sec:arc:stitches}

The GarmentCode approach to stitch representation is one of the key elements that enable modular component construction. We aim to keep the stitching abstracted from the internal structure of individual components and rather reflect a semantic connection between high-level components. This allows substituting one component in the connection by another with the same semantic meaning despite differences in underlying geometry -- enabling the interchangeability property. 
This behavior is realized by defining an abstract stitch as a connection between component-defined interfaces instead of a connection between panels' edges, as in flat sewing pattern representations. 
For example, in composite garments, the bottom of a bodice connects to the top of an under-waist garment. The same abstract stitch would describe a connection of a fitted bodice to a flared skirt, as in a 1950s dress, or a basic straight bodice to pants in a jumpsuit {(see \cref{fig:body_retargeting} for patterns of these examples)}. 

\vspace{-1mm}
\subsubsection{Interface}
\label{sec:arc:interfaces}

An interface describes how and where a particular component can be connected with. An interface contains a collection of edges of panels that can connect to another component in an abstract stitch. An interface can be constructed directly as a subset of panel edges (usually in panel components), or from reusing or combining interfaces of subcomponents. Thus, a single interface can contain multiple edges from multiple different panels (in contrast to 1-1 edge stitches in~\cite{Korosteleva2021GeneratingPatterns}). One component may have several interfaces.

In addition to stitches, interfaces may be useful for other purposes. For example, in sleeves, an interface specifies a projecting shape for correct modification of the bodice panel, which differs from the shape of the sleeve panel edges themselves. In~\cref{fig:garmentcomponents}, an Armhole is part of a Sleeve and defines the projecting shape, and \cref{fig:sleeves_shape} shows the differences and the projection result up close.

\vspace{-1mm}
\subsubsection{Stitching rule}

In our implementation, an abstract stitch is specified simply as a pair of interfaces, wrapped in a stitching rule object. The wrapper encapsulates the processing of the stitch flattening (see below), performed at stitch declaration time. 

\vspace{-1mm}
\subsubsection{Flattening stitch representation}
While abstract stitches are convenient for modeling, downstream processing tasks like simulation usually require a flat representation of stitches as edge-to-edge connection instructions. 
Unpacking the hierarchy of interfaces is straightforward, however, oftentimes one or both interfaces participating in a stitch contain multiple edges. Breaking such a stitch down to edge-to-edge instructions requires additional processing.
To perform this conversion, GarmentCode automatically generates additional vertices on the underlying panels to match the number of edges in the two connecting interfaces. Once the subdivision is completed, the set of resulting one-to-one stitches can then be used as a flattened representation. At the current stage of development, we use a simplifying assumption that each edge participates in no more than one stitch.

\input{fig_tex/vert_proj.tex}%

The process of generating the needed vertices is as follows. First, we note that the total length of edges in the interfaces on either side may not match, e.g., in stitches with gather, as in the Regency dress in~\cref{fig:historical_header}. To accommodate for non-matching lengths of interfaces, the edge lengths are represented as fractions of the total lengths of an interface instead of being used directly. The fractions from one of the interfaces are projected onto another interface to generate additional vertices whenever the projections fall outside of existing ones {with some tolerance}, and then the process is repeated in the other direction. The edge sequences in the interfaces are assumed to be aligned in the expected connection order.

Our stitch flattening algorithm is rather straightforward, and similar algorithms are likely to be behind many-to-many stitch features in commercial visual CAD like CLO3D~\cite{clo3d}.

\vspace{-1mm}
\subsection{Serialization}

Serialization denotes a conversion of a GarmentCode hierarchical component into a flat sewing pattern representation that can then be passed on to downstream tasks such as cloth simulation. {The process} is fairly straightforward: GarmentCode recursively converts all panels involved in component construction into a text representation and then gathers them into one file, {together with flattened stitching instructions.} 
In this work, we serialize component instances into the open-source JSON file format introduced in~\cite{Korosteleva2021GeneratingPatterns}, compatible with their draping pipeline. 

\vspace{-1mm}
\subsection{Parameterization format}

While the opportunity for defining parameters emerges naturally from using a programming-based paradigm, GarementCode provides configuration formats to support this feature further. In GarmentCode, we propose separating body and style parameters into two sets, with body parameters containing the measurements of the current avatar, and style parameters specifying not only the particular values but also the possible value ranges and type (numerical, boolean, or categorical) for each parameter, to allow for design sampling. 

Since some of the useful body measurements can be derived from others, GarmentCode provides an abstract class for loading body configuration files, with a separate method for the specification of formulas for derived parameters, which can be implemented in the application through a subclass. For example, in our prototype garment configurator, we use the waist level from the ground for positioning the garments, which is calculated from the body height, height of the head, and the usual waistline measurement from the nape of the neck to the waist. The calculation of derived design parameters, as well as the handling of parameter value conflicts, are usually component-specific and hence should be defined by the creators at the appropriate level.

\vspace{-1mm}
\subsection{Helpers}
\label{sec:arc:helpers}
We propose severalroutines designed to simplify the construction of sewing patterns. It is worth noting that the presented algorithms, when needed, were chosen for their simplicity while providing reasonable quality, however, better or more efficient solutions are likely to exist. All calculations are performed at evaluation time.

\vspace{-1mm}
\subsubsection{Typical edge sequences}
\label{sec:help:edge_seq}

We implement shortcuts to create typical edge sequences, which include creating a loop of straight edges from a set of vertices, edge subdivision {for all edge types} (adding vertices inside an existing edge specified by relative distances), and creating a triangular dart shape, specified by desired dart width and depth.

\vspace{-1mm}
\subsubsection{Defining curve edges}

In practice, curved edge design is often driven by certain requirements, rather than by the placement of control points. GarmentCode implements two optimization-based routines to specify Bézier curves: 
\begin{tightitemize}
\item Specifying a quadratic Bézier curve with the highest point at a particular location, used e.g.\ in pencil skirts and pants to correctly hug the hips in a manner transferable across different bodies;
    \item Smoothly adjusting a cubic Bézier curve to match the desired tangent directions at the edge ends while preserving the curve lengths, used in defining inverted sleeve opening shapes for curve-based sleeves. 
\end{tightitemize}
The second routine is described in~\cref{app:sec:curved_elements} as part of the sleeve inversion algorithm.

\vspace{-1mm}
\subsubsection{Projection operators}
\label{sec:help:projection}

\input{fig_tex/proj_operator.tex}%

Oftentimes when there is a need to connect two panels together, we wish the interface on one panel to be in stylistic or functional correspondence with the geometry of the other, e.g.\ when connecting a sleeve panel with a bodice panel. Describing and maintaining such correspondence across independent panels and components would break the encapsulation principle, so instead GarmentCode offers projection operators to transfer the shape defined in a one-panel component onto another panel. The transferred shape may be part of a panel geometry, or be a \emph{construction geometry} defined specifically for use in projection, implemented as an interface. Such a solution ensures that either of the panels can be easily modified or substituted, without the need to manually accommodate the changes in the second panel, supporting our target plug-and-play garment construction style. An example use of projections is demonstrated in~\cref{fig:construction}.

\input{fig_tex/garment_components.tex}

GarmentCode supports projection on an edge (injection) and projection on a corner for an arbitrary open chained edge sequence as projection shape, and edge or pair of chained edges, correspondingly, as target shapes. Both types support the use of curves in projection shapes, as well as in the target edges, and hence are formulated as optimization problems. 

In both cases, the goal is to find the points on the target that align with the ``opening'' (the first and the last vertices in the edge sequence) of projection shapes. The target edges are then split at the found points, and the projection shape is injected in between the found points into the target edge sequence, while the leftover ``cuts'' are removed. What differs between the two projection types is the process of finding the injection points. 
Projection on a corner relies on finding the points in the curve parameterization space (for straight edges and circular arcs we use arc length parameterization) whose 2D positions correspond to the projection shape opening:
\begin{multline}
    \argmin_{t_1, t_2}\left \| (e_1(t_1) - e_2(t_2)) - \textit{proj\_vec})  \right \|^2,  \\
    \text{s.t. } 0 \leq t_1 \leq 1,\ 0 \leq t_2 \leq 1,
\end{multline}   
where $e_1, e_2$ are the edges of the target corner, $t_1, t_2$ are values in curve parameterization space, and $\textit{proj\_vec} = \textit{proj.end} - \textit{proj.start}$ is the vector describing the projecting shape opening. 

Projection on edges also acts in curve parameterization space and finds two points that accommodate the projecting shape and are equidistant from the target point of injection $e(t)$: 
\begin{multline}
    \argmin_{t_1, t_2}(\left \| e(t + t_1) - e(t - t_2)\right \| -\left \| \textit{proj\_vec}  \right \|)^2  + \\
    + (\left \| e(t + t_1) - e(t) \right \| - \left \| e(t - t_2) - e(t) \right \|)^2, \\
    \text{s.t. } 0 \leq t_1 \leq 1,\ 0 \leq t_2 \leq 1,
\end{multline}   
where $e$ is the target edge, $t$ is the requested placement of the projecting shape $t_1, t_2$ are the shifts, all specified in respective curve parameterizations, and $\textit{proj\_vec} = \textit{proj.end} - \textit{proj.start}$ is the vector describing projecting shape opening. 

In case of projection on an edge, the projecting shape is automatically rotated such that $\textit{proj\_vec}$ aligns with the estimated insertion vector $(e(t + t_1) - e(t - t_2))$, following the edge direction. The inserted shape may be reflected over the insertion vector following a user-specified parameter to appear on the other side. Rotation alignment in projection on the corner is left to be specified externally to ensure design flexibility: within certain limits, different rotations of the same shape projected on a corner would produce different but equally valid results. 

\subsubsection{Smart copy}
\label{sec:help:smart_copy}

The garment designs often exhibit reflection symmetry w.r.t.\ the sagittal plane (left-right symmetry), so describing only one of the halves often suffices. To support such a design shortcut, GarmentCode provides a mirroring operator, which reflects a component over a vertical line, including its shape and the location (used in our upper garment components, showcased throughout the paper). For other types of repetitive designs, GarmentCode provides a distribution of components copied along a line or a circle, an example use of which is demonstrated in~\cref{fig:code}. 

\vspace{-1mm}
\subsubsection{Panel normal}

The notion of the right and wrong sides of a fabric is represented as the direction of the panel normal (positive orientation corresponding to the ``right'', usually outwards-facing side). The normal direction is defined by the counterclockwise traversal of edges. GarmentCode can automatically update the edge loop traversal such that the right side of the fabric points outside of the body, under the assumption that a body model is aligned with the axes in the coordinate basis in which the panel's 6D placement is specified. To do so, we evaluate the position of the panel center-of-mass (COM), and for each edge determine whether it traverses the COM on the right or on the left using the cross product of the edge vector from its end to start vertex with the vector from COM to the edge start vertex. The normal is then the prevalent direction of those cross products among the panel edges. If the normal is not oriented outwards of the center of the body, the edge loop is reversed to flip the normal. Large curvature arcs in curve edges may interfere with this process, so we use their linear approximation. We find the extremal points of the curve (furthest away from the straight line of an edge) and use them as new panel vertices, connecting them by straight edges.

In our implementation, the panel normal is adjusted upon every update to the panel's placement, ensuring correct normals throughout. However, updating the normals of all involved panels at the time of component serialization may also suffice.

\vspace{-1mm}
\subsubsection{Placement support}

GarmentCode simplifies the task of correctly manipulating the component placement. First, all placement modifications performed at component level are automatically propagated to subcomponents while preserving their relative placement (that was set on the lower levels of the hierarchy). Second, GarmentCode provides a helper to adjust the translation of one component so that its chosen interface is aligned with another component's interface in 3D, which can be used to align components by their stitches. The helper simply evaluates the 3D centers-of-mass of the edge sequences described by the interfaces, and the modified component's translation is updated by the location difference on the two COMs. The modified component may additionally be shifted outward of the \emph{component} COM to create a gap between two interfaces.

%% file: fig_tex/construction.tex
\begin{figure*}[ht]
  \centering
  \includegraphics[width=\linewidth]{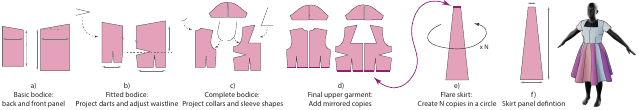}
  \caption{Construction of a fitted bodice component and a skirt for 1950s dress style pattern. Dashed arrows denote projection operators; gray arrows show some of the body-related and stylistic parameters of presented components; thick lines on d) and e) show the interfaces of upper garment and skirt components. 
  }
  \label{fig:construction}
\end{figure*}

%% file: fig_tex/arc_overview.tex
\begin{figure}[t]
  \centering
  \includegraphics[width=\linewidth]{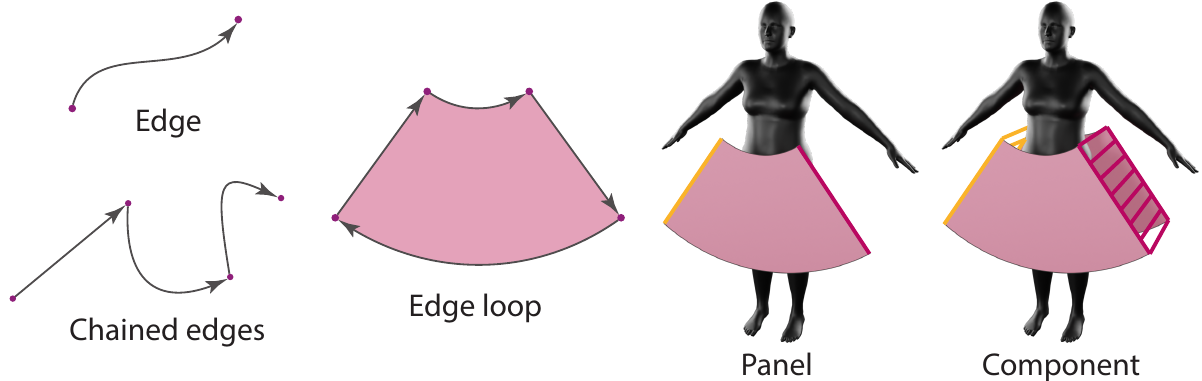}
  \caption{Overview of the elements of a GarmentCode architecture. Highlighted edges on panels correspond to chosen interfaces. The body model serves as a positioning reference. Here and in the figures below we use SMPL female average body model~\cite{Loper2015}, unless otherwise specified.
  }
  \label{fig:architecture}
  \vspace{-.6cm}
\end{figure}

%% file: fig_tex/vert_proj.tex
\setlength{\columnsep}{5pt}%
\setlength{\intextsep}{0pt}%
\begin{wrapfigure}{r}{0.3\linewidth}%
  \includegraphics[width=\linewidth]{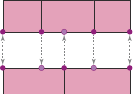}%
  \label{fig:vert_proj}%
\end{wrapfigure}%

%% file: fig_tex/proj_operator.tex
\setlength{\columnsep}{5pt}%
\setlength{\intextsep}{0pt}%
\begin{wrapfigure}{r}{0.55\linewidth}%
  \includegraphics[width=\linewidth]{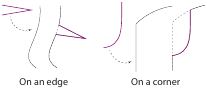}%
  \label{fig:proj_op}%
\end{wrapfigure}%

%% file: fig_tex/garment_components.tex
\begin{figure*}[t]
  \centering
  \includegraphics[width=0.9\linewidth]{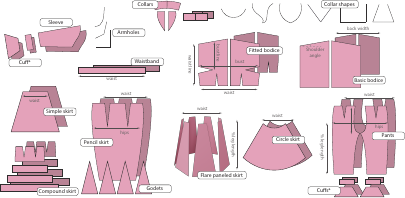}
  \caption{Samples of garment components that we design using GarmentCode (stitches are not depicted for clarity). Each component's appearance is conditioned on style parameters and body measurements, some of which are illustrated with gray arrows. *\textit{Cuffs for sleeves and pants are the same component, shown twice to demonstrate potential use in both cases.}
  }
  \label{fig:garmentcomponents}
\end{figure*}

%% file: sec/4_application_garments.tex
\section{Application: Garment configurator}
\label{sec:configurator}

We apply the GarmentCode architecture described in \cref{sec:architecture} to build a garment configurator. The configurator allows selecting and assembling various high-level components and interactively manipulating their parameters, displaying the resulting sewing pattern. We design a collection of parametric garment components: various styles of skirts, such as flare, godet, pencil, gather, and compound, {with the flare skirt implemented with two different topologies};  bodice (i.e., components covering the torso, which can be fitted or loose), pants, sleeves (with optional cuffs), and different collar shapes. See \cref{fig:garmentcomponents}. Many of these components can represent garments on their own, but an additional meta-component enables combining those elements into complex dresses (\cref{fig:historical_header}) and jumpsuits (\cref{fig:body_retargeting}). Components are parameterized w.r.t.\ body measurements (bust size, waist length, etc.), and style features (like lengths of elements). Most of the style parameters are defined to depend on body measurements or each other (e.g., collar width is bound between neck width and shoulder size), which additionally enables body retargeting and ensures pattern validity. 

A selection of garments sampled from our parametric template is presented in~\cref{fig:designsamples}. An example garment program is provided in~\cref{fig:code}. Flexible implementation of our GUI allows any garment programs with the recommended parameterization format to be loaded as GUI with little-to-no tweaks.

\input{fig_tex/sleeve_shapes.tex}
\input{fig_tex/sleeves.tex}

\subsection{Example construction process}

Here we describe an example process of constructing a dress in a 1950s style with GarmentCode. This dress requires a definition of a flare skirt, fitted bodice, collars, and sleeves. The process is illustrated in~\cref{fig:construction}. 

A fitted bodice is created to accommodate the natural body curvature and accentuate the waistline; darts are employed to create this effect. Since we assume that the body is left-right symmetric, we start building this piece by defining two quadrilateral panels representing one half of the front and the back of the bodice, following the body measurements {(\cref{fig:construction}, a)}. The front panel is wider and longer to accommodate the extra curvature on the chest. We add darts {(\cref{fig:construction}, b)} by projecting a triangular shape onto the side of the front panel, such that the side with a dart is the same length as the back side, and the bottom of the front and back panel, and removing some length from the side as well. The bottom length of both pieces equals half of the body waist measurement. 

We combine the bodice component with other components to create an upper garment. First, we define a collar shape and project it onto the inner corners of the front and back panels. Secondly, we take a sleeve component and project a corresponding opening shape onto the outside corners of the front and back panels (\cref{fig:construction}, c; \cref{fig:sleeves_shape}). 
Applying the projection operator makes the bodice design agnostic to the types of sleeves and collars used in this step, hence these components are easily replaceable with other designs. 
The next step is simply to mirror the upper garment component created so far, as described in~\cref{sec:help:smart_copy}. The bottom edges of the final four panels are designated as the interface of this upper component {(\cref{fig:construction}, d)}.

\input{fig_tex/body_retargeting.tex}

We approximate a 1950s-style flare skirt with trapezoid panels replicated multiple times and distributed around the body {(\cref{fig:construction}, e)}. { The same shape can be achieved with two or even one panel -- a section of a circle, but we show a more complex option to demonstrate the capabilities of GarmentCode.} The tops of the panels should follow the body waist size, and the bottoms are wider, creating a flare effect. The sequence of the top edges of the skirt panels is designated to be a skirt interface. 
The final step is to connect the bottom of the upper garment with the top of the skirt, which is done automatically (see~\cref{sec:arc:stitches}).  

\section{Evaluation}
\label{sec:evaluation}

\subsection{Element parameterization}

The power of parameterization in garment design can be well demonstrated on a sleeve example~\cref{fig:sleeves}. The same building block can produce various sleeve silhouettes, from modern streetwear style to a vintage balloon sleeve, merely by varying a few parameters. The GarmentCode representation and parameterization open up an easy way for experimenting with designs based on human-readable parameters, rather than editing sewing patterns at the low level.

{
\subsection{Handling complexity}
\label{sec:eval:complexity}

\input{fig_tex/compound_skirt}

The convenience of manipulating garments through hierarchical component structure is well-demonstrated through an example of a compound skirt (\cref{fig:compound_skirt}). Having the base skirts (pencil, flare, gather) and their corresponding parameter spaces defined, creating a compound skirt that uses existing components as layers is trivial: it requires an initialization of the base skirt (hugging the hips), and multiple copies of the skirt type used for the different levels, initializing their size based the bottom size of the one above, connecting their tops to bottoms of previous ones, and using the placing helper for correct alignment. With the base skirts implemented, extending our component library with the compound skirt component takes just 50 lines of code (see \textsc{skirt\_levels.py} in supplementary code). Thanks to the component abstraction, it seamlessly integrates into the library: it can be used in place of other skirts or pants by simply adding its class name to the list of supported components in the meta-component parameter range for bottoms types.

\subsection{Body retargeting}

Our garment components are parameterized by body measurements, which makes it easy to fit a garment design on a different body. The parameterizations are introduced in the areas of garments that tightly hug the body, e.g., the waist of flare skirts and pants. The fitted bodice component is fully specified by body measurements (waist and bust circumferences, bust line, waistline, and back width) since its purpose is to accentuate the body curves. Some style parameters are made dependent on the body measurements, e.g., the length of a flare skirt is specified as a fraction of leg length and varies according to the wearer's height.

To demonstrate body retargeting, we take several body shape samples from SMPL~\cite{Loper2015} and manually acquire their body measurements. \cref{fig:body_retargeting} shows the retargeting results. A 1950s-style dress recognizable hourglass silhouette relies on a proper fit of the tailored bodice component. 
The correct fit of the dress is fully preserved across large body shape variations thanks to the semantic encoding of GarmentCode components. The skirt length varies with the leg length, hence leaving approximately the same part of the leg uncovered in different body shapes. 
The tight-fitting pants in the jumpsuit and the pencil skirt in the strapless dress successfully adapt to different body shapes and proportions. The bottom of the pencil skirt is parameterized relative to the hip size, enabling the preservation of the defining upside-down triangular silhouette across all body models.}

\input{fig_tex/real_garment}

\subsection{Reproducing a real-world pattern}
\label{sec:eval:real_pattern}

To evaluate the patterns created with GarmentCode, we reproduce one of the professional garment patterns from Mood Fabrics~\shortcite{MoodFabrics:BirchDress} by adjusting the parameters of the demo configurator, see \cref{fig:real_garment}. We observe that the overall 3D shape and design intention are well reproduced. However, a number of details vary. Our panel definition excludes inner loops such as diamond darts, resulting in an excessive stitch in the waist area. Without the support for fold lines (both in our system and in the downstream simulator), the sleeve panel needs to be defined as two. Since we currently do not support stitching multiple layers of fabric, we can only model one-sided cuff and turtle neck components, whereas in the original pattern these are double-sided, with half of the panel folded inside.

Other differences are not dictated by the architectural limitations, but simply by the choices made when designing our example garment components. Our sleeve element uses smoother curves compared to the original pattern, resulting in a mismatch. The difference in lengths between the waist and the hip is distributed between the darts and the skirt sides differently, resulting in misaligned darts. Remaining variations (e.g.\ side dart width) are simply due to differences in body sizes between our body model and the standard sizing used in the original pattern.
The showcased limitations provide inspiration for further development of the GarmentCode system.

%% file: fig_tex/sleeve_shapes.tex
\begin{figure}[t]
  \includegraphics[width=0.8\linewidth]{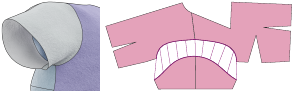} 
  \caption{Connecting a sleeve to a bodice. The edges on the sleeve side should create a concave shape compared to the sleeve opening shape on the bodice, so that the sleeve curves away from the body when stitched (see details in~\cref{app:sec:curved_elements}). Our sleeve component defines a bodice opening shape as an additional interface for correct projection and connection.
  }
  \label{fig:sleeves_shape}
  \vspace{-0.3cm}
\end{figure}

%% file: fig_tex/sleeves.tex
\begin{figure}[t]  
  \centering
  \begin{overpic}[trim=0cm 0cm 0cm -2.5cm,clip,width=1\linewidth,grid=false]{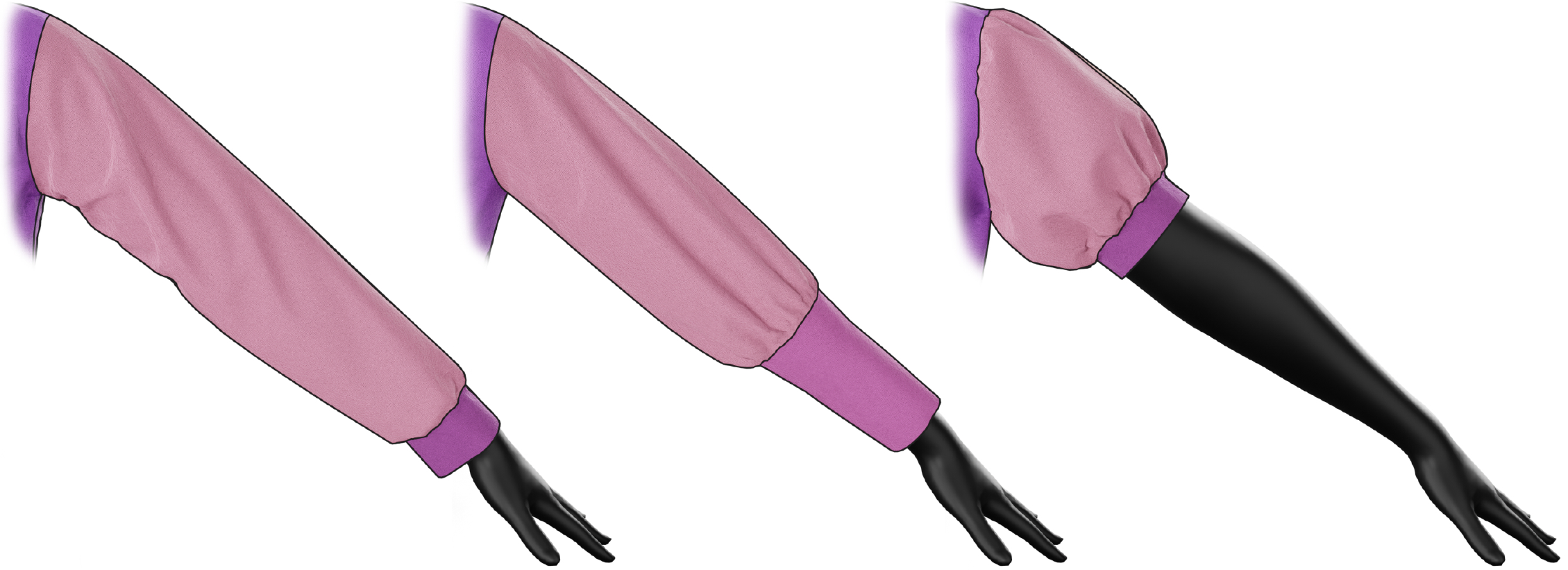}
  \end{overpic}
  \caption{Sleeve styles created from the same sleeve component by varying sleeve length, cuff length, and gather parameters
  }
  \label{fig:sleeves}
  \vspace{-0.4cm}
\end{figure}

%% file: fig_tex/body_retargeting.tex
\begin{figure*}[t]
  \centering
  \includegraphics[width=0.95\linewidth]{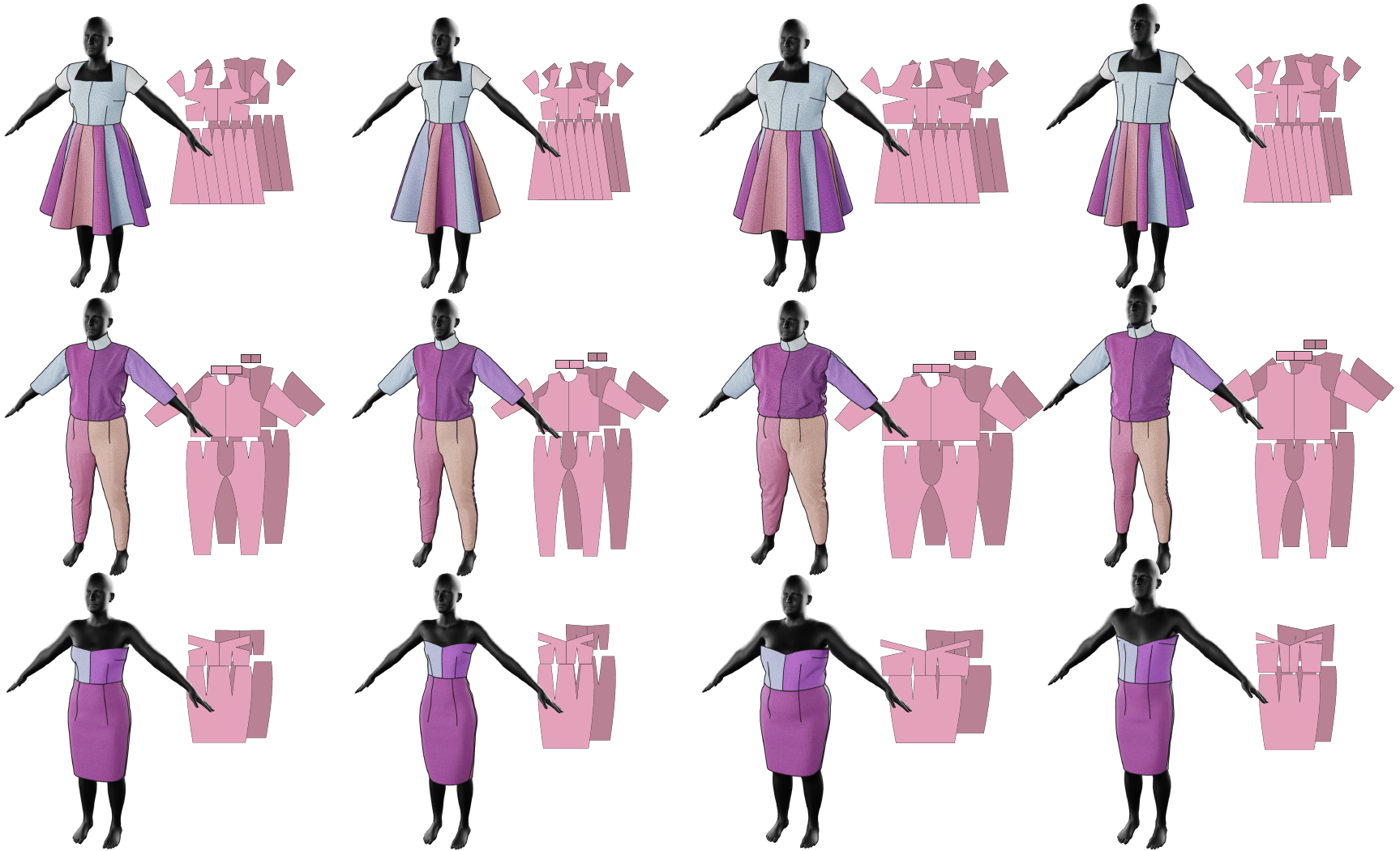}
  \caption{Retargeting garments conditioned on body measurements across different body shapes. 
  }
  \label{fig:body_retargeting}
\end{figure*}

%% file: fig_tex/compound_skirt.tex
\begin{figure}[tb]
  \centering
  \includegraphics[width=\linewidth]{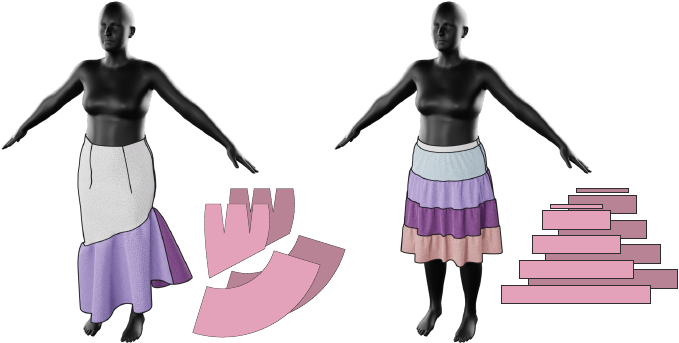}
  \caption{Two design samples from a compound skirt component. With trivial implementation, this component allows combinations of existing skirts and their parametrizations, resulting in complex garment styles.}
  \label{fig:compound_skirt}
\end{figure}

%% file: fig_tex/real_garment.tex
\begin{figure}[tb]
  \centering
  \includegraphics[width=\linewidth]{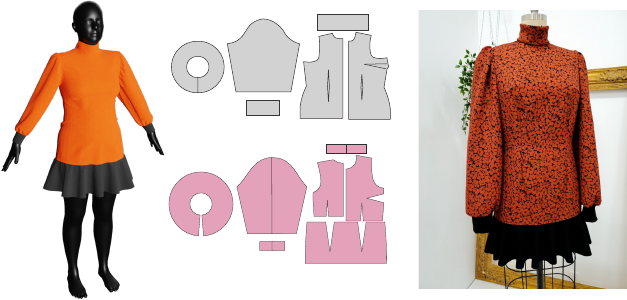}
  \caption{Reproducing a production sewing pattern ``Birch dress'' of MoodFabrics~\shortcite{MoodFabrics:BirchDress}. The MoodFabrics pattern is in grey (top) with the final garment on the right (photo provided by MoodFabrics), and ours is in pink (bottom) with the final garment on the left. Some discrepancies occur simply due to different design decisions and body sizes, while others highlight the limitations of the base GarmentCode architecture, as discussed in~\cref{sec:eval:real_pattern}.
  }
  \label{fig:real_garment}
  \vspace{-.4cm}
\end{figure}

%% file: sec/5_discussion.tex
\section{Discussion}
\label{sec:discussion}

We introduced a new framework for representing and designing parameterized garments. Our architecture encourages composing garments as hierarchical structures with interchangeable parametric components akin to configurable puzzle pieces. This approach enables exponential growth of design possibilities whenever a new component is added to the collection, expanding the design space, which can be easily explored through semantic parameters with little manual overhead, or sampled when constructing design datasets.  

We demonstrated how our framework can be employed to create parametric garment design templates suitable for product configurators or in design samplers for synthetic garment datasets. Our templates offer extended design spaces, garment transfer across different body shapes, and produce valid sewing patterns for each instance, which can be passed on to a physics-based simulator or adapted for fabrication.

Our presented system changes the paradigm of garment construction to programming-based, which does not follow the traditional design workflow, and presents other challenges like the need for explicit specification of vertex coordinates in panels, which may be an obstacle for industry adoption. However, the successful cases of embracing programming in the creative domains, such as solid modeling CAD/DSL systems~\cite{Featurescript,openscad}, procedural tools for plans~\cite{Makowski2019}, buildings~\cite{Muller2006ProceduralBuildings}, or city landscapes~\cite{Parish2001ProceduralCities}, and even such widespread fields as web design, give us reason to believe that fashion creators might be willing to acquire the needed programming skills to access unique features GarmenCode provides. Pairing programming-based parametric construction with a visual tool for specifying panel and edge geometry could be an interesting avenue for future work.

Creating a new design tool for garment construction is an ambitious and complex goal. GarmentCode aims to demonstrate the potential of our idea and provides a solid proof-of-concept implementation, but it is not all-encompassing. GarmentCode could be expanded with additional helpers to improve the toolkit: readjusting the edge shape after dart insertion for a smooth connection, adding a curved dart calculator, adding rotation alignment to the placement by stitches, etc. On the architecture level, the simplified definition of a panel does not allow specification of internal loops, hence GarmentCode cannot seamlessly represent panels with holes (\cref{fig:real_garment}). The architecture could also be extended to incorporate elements that are sewn on top of a fabric piece, such as pockets and flounces. Likewise, GarmentCode currently has limited support for sharp folds, and more tools are needed to efficiently specify and assemble pleats and smocking patterns. We also wish to further accommodate the differences between sewing patterns for garment fabrication vs.\ simulation: for example, merging excessively fragmented panels in the final pattern to reduce the number of stitches needed (e.g., removing the central stitch for front and back panels). Representing different stitch appearances in the spirit of~\cite{Rodriguez2022TrueGarments} is also an interesting direction to explore. 

Finally, there is a considerable variation in sewing pattern geometry even within the basic garment elements, which is not fully represented in our implemented garment components, making it difficult to reproduce real garment designs merely by varying the semantic parameters of the demo configurator (\cref{fig:real_garment}). An additional engineering effort is required to accommodate such variations of real-world patterns.

GarmentCode achieves several considerable advancements through a simple architecture. It is evident that the problem of garment construction is under-explored, and we hope our work will inspire further research on computational support of this important engineering problem.

%% file: sec/6_acks.tex
\begin{acks}

This work was partially supported by the European Research Council (ERC) under the European Union’s Horizon 2020 Research and Innovation Programme (ERC Consolidator Grant, agreement No.\ 101003104, MYCLOTH). Autodesk and Qualoth provided licenses for their software. We thank Anna Hrustaleva for the invaluable consultations on fashion styles, and the members of IGL for always coming to the rescue in times of need for discussion and sweets.

\end{acks}

%% file: fig_tex/random_design_samples.tex
\begin{figure*}[ht]
  \centering
  \includegraphics[width=\linewidth]{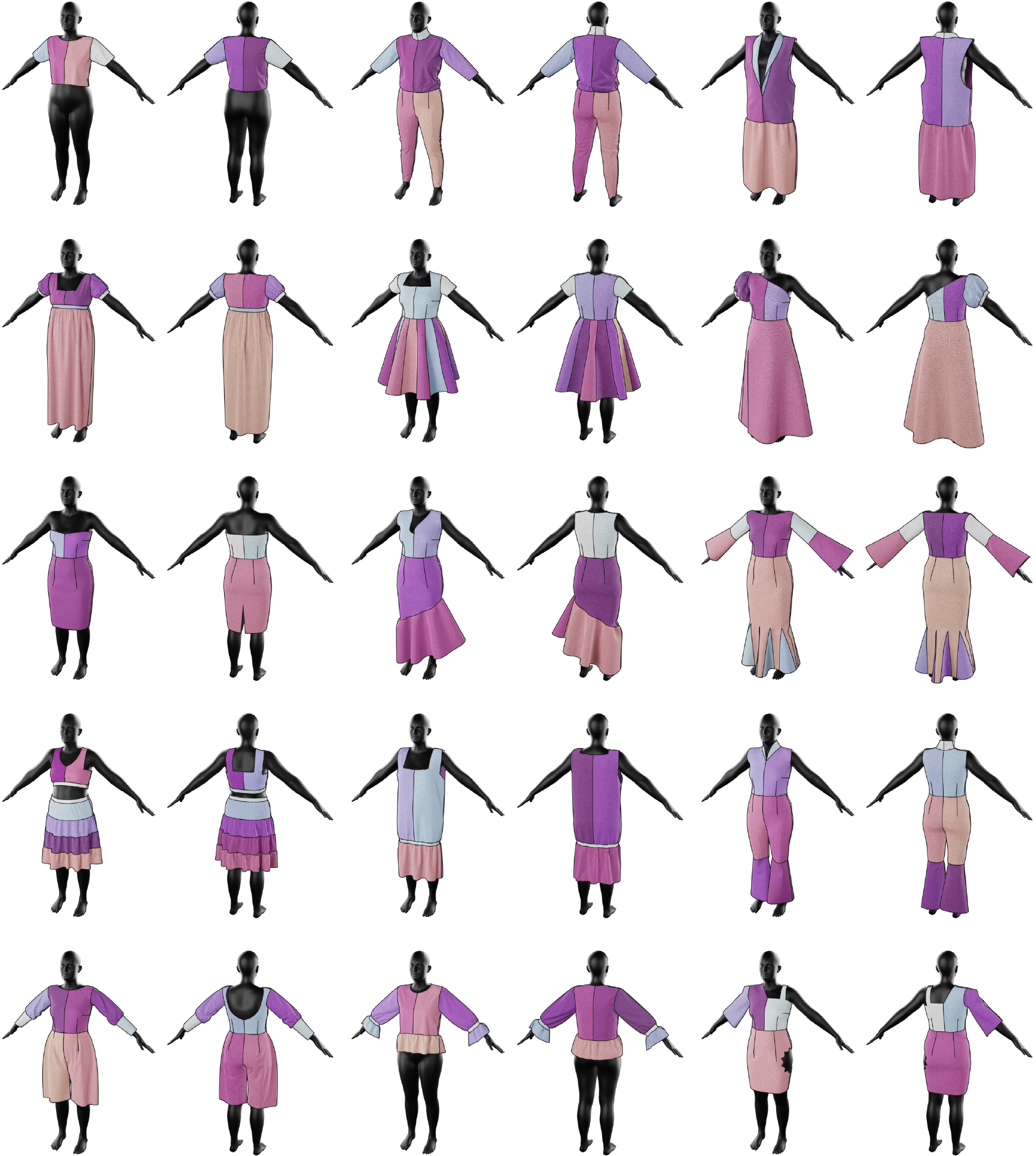}
  \caption{A selection of design samples from our parametric garment template. The segmentation corresponds to panels and stitches in the respective sewing patterns. 
  }
  \label{fig:designsamples}
\end{figure*}

%% file: fig_tex/code.tex

\begin{figure*}[ht]
  \centering
  \includegraphics[width=\linewidth]{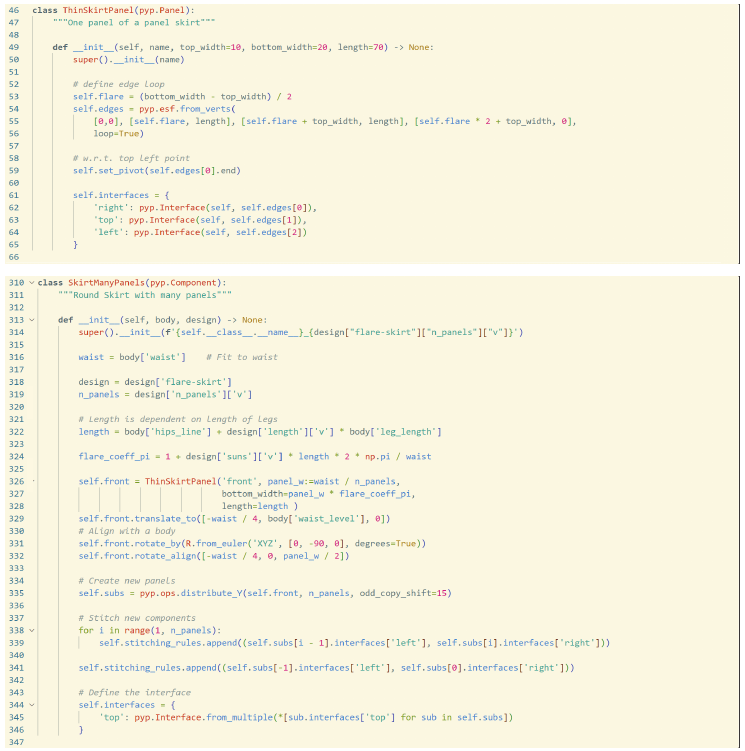}
  \caption{Example of a garment program written with GarmentCode, showcasing a skirt used in our 1950s dress example. The \textsc{ThinSkirtPanel} class defines a single trapezoid-shaped panel. \textsc{SkirtManyPanels} creates an instance of the panel according to style parameters (desired length, number of panels, and flare, where 1 sun = full circle skirt), and body measurements (waist to condition the top opening, and hips height to condition the length). The panel is then placed, and its copies are distributed on a circle around the body (using \textsc{distribute\_Y()} operator) and stitched. The top edges of all panels constitute the interface of a final skirt.
  }
  \label{fig:code}
\end{figure*}

%% file: sec/Supplementary.tex
\clearpage
\appendix

\section{Constructing curved elements}
\label{app:sec:curved_elements}

One of the advantages of the programming-based paradigm of garment modeling is the ability to utilize computational tools for defining garment elements that are difficult to specify correctly by hand, e.g., when it comes to manipulating smooth curves. Here we elaborate on one such example to complement the experiments presented in the paper. 

\subsection{Inverting sleeve opening}

As shown in~\cref{fig:sleeves_shape}, the connection between the sleeve and a chosen bodice block is non-trivial, requiring the shape of the sleeve to be an ``inverse'' of the shape of the sleeve opening on the bodice in order to correctly wrap around the arm. In addition to these constraints, the shape of the inverse connection on the sleeve defines the rest angle of the sleeve in 3D: the arm angle at which there are neither folds nor tensions in the garment fabric. Smaller angles allow putting arms up more easily and are thus good for activewear, while bigger angles create fewer folds in the armpit zone when the arms are down, hence more suitable for officewear, as shown in~\cref{fig:supp:sleeve_rest_shape}.  

For sleeve openings based on curves, defining sleeve edges correctly is especially challenging, since the inversion should preserve the length of the edge while following the desired rest angle and maintaining smoothness of connection of sleeve panels. GarmentCode helps with this task.

Our process assumes that the projecting shape for a bodice is defined as a cubic Bézier curve with both control points on one side of the edge. The first step is then to create an initial guess for the inverted sleeve shape: the control point towards the end of the edge is flipped to the other side (flipping the $y$-coordinate in its relative representation), and the edge direction is aligned with an axis perpendicular to zero angle sleeve rest shape. In our implementation, the $x$-axis corresponds to a fully horizontal sleeve, and the perpendicular is the vertical direction. The edge is then rotated by a desired rest angle~$\theta$. 

\input{fig_tex/supp_sleeve_notation}
The second step is an optimization process, in which the edge extension and new positions of
curvature control points {are optimized} s.t.\ the length of the curve is preserved, while the curve tangents at the endpoints are aligned with the downward direction at the top and the desired sleeve angle at the bottom. The first condition ensures a smooth connection at the top between the front and back sleeve panels, while the second one enables the inversion effect and supports the chosen rest angle. The function to minimize is as follows: 
\begin{multline*}
   E(c_1, c_2, s) = (\left \| e(c_\text{start}, c_1, c_2, c_\text{end} + s \cdot v) \right \| - l)^2 + \\    
        + \left \|T_0(e(c_\text{start}, c_1, c_2, c_\text{end} + s \cdot v)) - T_0^*\right \|^2 + \\    
        + \left \|T_1(e(c_\text{start}, c_1, c_2, c_\text{end} + s \cdot v)) - T_1^*\right \|^2 + \\    
        + \lambda\left( C_\mathrm{max}\left(e\left(c_\text{start}, c_1, c_2, c_\text{end} + s \cdot v\right)\right) \right)^2,
\end{multline*}   
where $c_1, c_2$ are the cubic Bézier control points, $c_\text{start}, c_\text{end}$ are initial edge endpoints, $e(c_\text{start}, c_1, c_2, c_\text{end})$ is a curved edge with given endpoints and control points, $s$ is the scaling factor of the edge vector $v = c_\text{end} - c_\text{start}$, the $T_0(\cdot),\ T_1(\cdot)$ functions evaluate the curve tangent at the start and the end point of the edge curve, with $T_0^*,\ T_1^*$ being the target tangent values as described above. Finally, $C_\mathrm{max}(\cdot)$ evaluates the maximum curvature of the edge and is used to regularize the curve smoothness. 

The given process produces a correct sleeve inversion for sleeve openings of arbitrary size and for a desired sleeve rest angle, allowing us to define both as garment style parameters, as is done on our prototype garment configurator. The optimization process is included in the core GarmentCode as an operator.

\input{fig_tex/sup_sleeve_rest}

%% file: fig_tex/supp_sleeve_notation.tex
\setlength{\columnsep}{5pt}%
\setlength{\intextsep}{0pt}%
\begin{wrapfigure}{r}{0.35\linewidth}%
  \includegraphics[width=\linewidth]{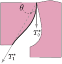}%
  \label{fig:supp:sleeve_notation}%
\end{wrapfigure}%

%% file: fig_tex/sup_sleeve_rest.tex
\begin{figure}[t]
  \includegraphics[width=\linewidth]{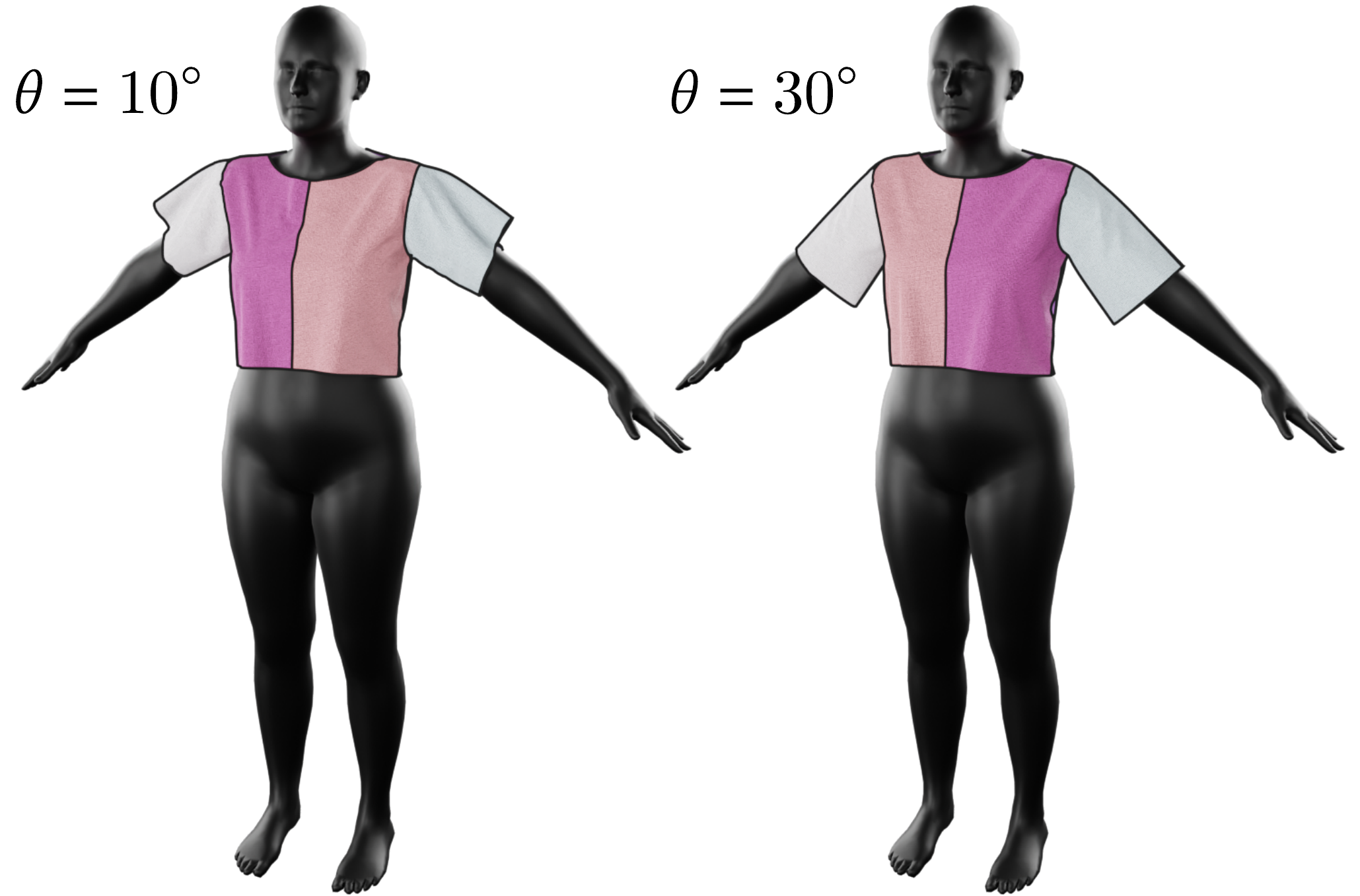} 
  \caption{T-shirts designed with different rest angles of sleeves (denoted $\theta$) draped in the same pose. Note the differences in how well the fabric follows the arm angle. 
  }
  \label{fig:supp:sleeve_rest_shape}
\end{figure}